\documentclass[prd ]{revtex4}%
\usepackage{caption}
\usepackage{amsmath}
\usepackage{graphicx}
\usepackage{epsfig}
\usepackage{dcolumn}
\usepackage{bm}
\usepackage{slashed}
\usepackage{amsfonts}
\usepackage{amssymb}%
\providecommand{\U}[1]{\protect\rule{.1in}{.1in}}
\begin{document}
\title{ Can the symmetry breaking in the SM \\be determined by the Top-Higgs Yukawa interaction?}
\author{Jose Carlos Suarez Cortina $^{*}$, Denys Arrebato $^{*}$ and  Alejandro
Cabo$^{**}$  \bigskip}
\address{$^{*}$  Instituto de Tecnolog\'ias y Ciencias Aplicadas, InStec, La Habana.\bigskip \\
$^{**}$ Department of  Theoretical Physics, Instituto de Cibern\'etica, Matem\'atica y F\'isica ICIMAF, La Habana, Cuba. \bigskip }

\begin{abstract}
\noindent In this letter, we first resume the results of a previous article
(EPJC(2011)71:1620).  That work considered a simple model of QCD including a
Yukawa interaction with a scalar field. Its two loop effective potential for
the scalar field, predicted a 126 GeV Higgs mass after the minimum of the
potential was fixed at a mean scalar field giving a 175 GeV Top quark mass.
However, a high value of the strong coupling was required ($\alpha=\frac
{g^{2}}{4 \pi}$ close to 1) to get these values. After reviewing the results
of this  study, an idea for extending the work is simply advanced here: to
consider the running strong coupling, in order to decide whether or not, the
usual values of the strong interactions have  the chance of justifying the
experimentally known values of the Higgs and Top quark masses. It is
underlined that a positive result of the proposed task will suggests the
possibility of basing the SM breaking of symmetry, on the so called
\textit{second minimum} of this model. This could also
identify the essential role of QCD in this effect. Results of the further
examination of this question will be presented elsewhere.

\end{abstract}
\maketitle

In ref. \cite{cabo} a simple massless QCD model including only one quark type
and a singlet scalar field with a Yukawa interaction between them, was
investigated. The aim of the study was to explore a suspicion about that the
so called "second minimum" of the Standard Model (SM) could in fact be
responsible for the symmetry breaking in the SM. As it is known, this minimum,
in addition to the usual one exhibited by the Higgs potential, is the result
of the Yukawa interaction of the Higgs field with the Top quark
\cite{second-1,second-2,second-3}. This idea emerged after noting that this
new minimum got its relevance only after the SM calculations arrived up to the
two loop level. Then, the question emerges about what could be the result of
an attempt to construct the SM around this new radiative corrections
determined minimum. Up to our knowledge, there had not been attempts to answer
this question in the past literature. These are the main motivations in
considering the study in the work in reference \cite{cabo}. The results of
that work were inconclusive, in spite of the fact the correct experimental
values of the Higgs and the Top quark masses were able to be fixed by choosing
a definite value of the strong coupling parameter. It happened, that the value
of this parameter was a high one: $\frac{g^{2}}{4\pi}$ close to $1$. In this letter we
start by reviewing the results in \cite{cabo} for afterwards simply propose an
idea which could perhaps justify to further study the possibility of basing
the SM in the so called "second minimum".\newline 

\begin{figure}[h]
\centering
\includegraphics[width=7cm]{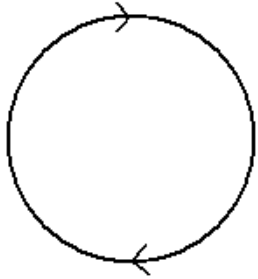} \caption{ The figure shows the quark one
loop correction. The result depends on the scalar \textquotedblright
mass\textquotedblright\ field $\phi$ through the quark free propagator which
is the usual free Green function of QCD, in which the mass is substituted by
$\phi$.}%
\label{fig1}%
\end{figure}

Let us now start reviewing
the main elements of the model discussed in \cite{cabo}. The generating
functional of the model is based in an action including a simple singlet
scalar field interacting with only one type of quark. The functional was
chosen in the form
\begin{equation}
Z[j,\eta,\overline{\eta},\xi,\overline{\xi},\rho]=\frac{1}{\mathcal{N}}%
\int\mathcal{D}[A,\overline{\Psi},\Psi,\overline{c},c,\phi]\exp[i\text{
}S[A,\overline{\Psi},\Psi,\overline{c},c,\phi]],\label{Z}%
\end{equation}
The action was taken in the form written below, in which in addition to the
usual massless QCD action, it was only considered a quark field Yukawa
interacting with a scalar field. To simplify the discussion, the free action
of the scalar field was defined as a massless free term in the absence of
self-interaction. The various terms in the action, after decomposed in free
and interaction parts, are written below
\begin{align}
S &  =\int dx(\mathcal{L}_{0}\mathcal{+L}_{1}\mathcal{)},\\
\mathcal{L}_{0} &  =\mathcal{L}^{g}+\mathcal{L}^{gh}+\mathcal{L}%
^{q}+\mathcal{L}^{\phi},\\
\mathcal{L}^{g} &  =-\frac{1}{4}(\partial_{\mu}A_{\nu}^{a}-\partial_{\nu
}A_{\mu}^{a})(\partial^{\mu}A^{a,\nu}-\partial^{\nu}A^{a,\mu})-\frac
{1}{2\alpha}(\partial_{\mu}A^{\mu,a})(\partial^{\nu}A_{\nu}^{a}),\\
\mathcal{L}^{gh} &  =(\partial^{\mu}\chi^{\ast a})\partial_{\mu}\chi
^{a},\label{S}\\
\mathcal{L}^{q} &  =\overline{\Psi}(i\gamma^{\mu}\partial_{\mu})\Psi,\\
\mathcal{L}^{\phi} &  =\partial^{\mu}\phi\partial_{\mu}\phi,
\end{align}

\begin{align}
\mathcal{L}_{1} &  =-\frac{g}{2}f^{abc}(\partial_{\mu}A_{\nu}^{a}%
-\partial_{\nu}A_{\mu}^{a})A^{b,\mu}A^{c,\nu}-g^{2}f^{abe}f^{cde}A_{\mu}%
^{a}A_{\nu}^{b}A^{c,\mu}A^{d,\nu}-\nonumber\\
&  -gf^{abc}(\partial^{\mu}\chi^{\ast a})\chi^{b}A_{\mu}^{c}+g\overline{\Psi
}T^{a}\gamma^{\mu}\Psi A_{\mu}^{a}+\overline{\Psi}\Psi\phi.
\end{align}
\newline After constructing the Feynman expansion being associated to the
above generating function and classical action, the evaluation of the
effective potential as a function of an homogeneous scalar (Higgs resembling)
field was considered in reference \cite{cabo}, up to the two loop
approximation. The one loop term of the potential as a function of the scalar
field $\phi$ was defined by the quark one loop diagram with the mean scalar
field as a background. It took the form:
\begin{align}
\Gamma^{(1)}[\phi]  & =-V^{(D)}N\int\frac{dp^{D}}{i\text{ }(2\pi)^{D}%
}Log[Det\text{ }(G_{ii^{\prime}}^{(0)rr^{\prime}}(\phi,p))],\\
D  & =4-2\epsilon.\nonumber
\end{align}
This contribution is illustrated in figure \ref{fig1}.  After evaluating the color and spinor traces and integrating in
dimensional regularization the effective action density took the form
\begin{equation}
\gamma^{(1)}[\phi]=\frac{\Gamma^{(1)}[\phi]}{\frac{V(D)}{\mu^{2\epsilon}}%
}=\frac{2N\text{ }\Gamma(\epsilon-1)}{(\frac{D}{2})(4\pi)^{2-\epsilon}}%
\phi^{2}(\frac{\phi}{\mu})^{-2\epsilon}.
\end{equation}
Next, by subtracting the pole in $\epsilon$ in order to employ the MS
substraction scheme, passing to the limit $D=4-2\epsilon\rightarrow$ 4 and
changing the sign, gave the Higgs potential one loop contribution in the form
\begin{equation}
v^{(1)}[\phi]=-\frac{3\phi^{4}}{8\pi^{2}}(-3+2\gamma+2Log(\frac{\phi^{2}}%
{2\pi\mu^{2}})).
\end{equation}
Note the important fact that the one loop potential for the scalar field
determined by the quark loop, is unbounded from below.\newline

\begin{figure}[h]
\centering
\includegraphics[width=7cm]{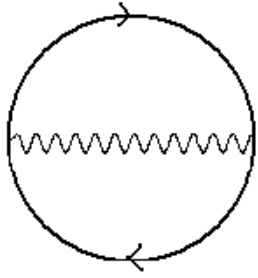}  \caption{ The two loop term determined
by the strong interaction. Again the $\phi$ dependence of the result is
introduced though the free quark propagator.}%
\label{fig2}%
\end{figure}

The two loop contribution associated to the quark gluon interaction in the
background of the scalar Higgs field in reference \cite{cabo} is illustrated
in figure \ref{fig2}. 

By writing the diagram expression and calculating the color and spinor traces,
this term was written in the form
\begin{equation}
\Gamma^{(2)}[\phi]=-V^{(D)}g^{2}(N^{2}-1)\int\frac{dp^{D}dq^{D}}{i^{2}\text{
}(2\pi)^{2D}}\frac{D\phi^{2}-(D-2)p.(p+q)}{q^{2}(p^{2}-\phi^{2})((p+q)^{2}%
-\phi^{2})}.
\end{equation}
The integrals appearing were integrated by employing the results in
\cite{fleischer}. After dividing by the volume the total action density,
associated to this two loop contribution, decomposes in the following two
dimensionally regularized terms:
\begin{align}
\gamma^{(2,1)}[\phi]  & =\frac{\Gamma^{(2,1)}[\phi]}{\frac{V(D)}%
{\mu^{2\epsilon}}}=-v^{(2,1)}[\phi]\nonumber\\
& =-\frac{2g_{0}^{2}(N^{2}-1)}{\text{ }(2\pi)^{2D}}\frac{A(\epsilon
)\pi^{4-2\epsilon}}{\epsilon^{2}}\phi^{4}(\frac{\phi^{2}}{\mu^{2}%
})^{-2\epsilon} \ \ \ \ \ \ \ \ \ \
\end{align}
and
\begin{align}
\gamma^{(2,2)}[\phi]  & =\frac{\Gamma^{(2,2)}[\phi]}{\frac{V(D)}%
{\mu^{2\epsilon}}}=-v^{(2,2)}[\phi]\nonumber\\
& =V^{(D)}\frac{g_{0}^{2}\mu^{2\epsilon}(N^{2}-1)2(1-\epsilon)}{\text{ }%
2(2\pi)^{8-4\epsilon}}\pi^{4-2\epsilon}(\Gamma(\epsilon-1))^{2}\phi^{4}%
(\phi^{2})^{-2\epsilon}.
\end{align}
Further, by minimally subtracting the poles of these expressions, taking the
limit $\epsilon\rightarrow0$ and numerically evaluating all the constants for
getting a simpler expression, the potential density (the negative of the
action density) associated to this two loop contribution becomes
\begin{align}
v^{(2)}[\phi]  & =5.34687\times10^{-5}\phi^{4}{\LARGE (}1227.16+474.276\text{
}g_{0}^{2}-\nonumber\\
& 12.0(\text{ }29.6088+13.5253\text{ }g_{0}^{2})\text{ }Log(\frac{\phi^{2}%
}{\mu^{2} })+\nonumber\\
& 12.0\text{ }g_{0}^{2}\text{ }(Log(\frac{\phi^{2}}{\mu^{2}}))^{2}{\LARGE ).}%
\end{align}
In this expression, the leading logarithm term in the scalar field is squared
and has a positive coefficient, making the result bounded from below.\newline

\begin{figure}[h]
\centering
\includegraphics[width=7cm]{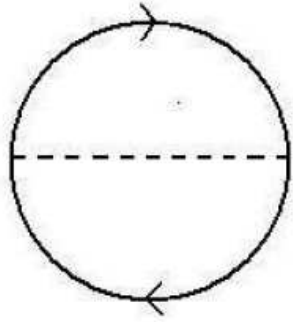}  \caption{ The diagram corresponding to
the radiative corrections of the ''mass'' field $\phi$ in the two loop order.
}%
\label{fig3}%
\end{figure}

Finally, in reference \cite{cabo} it was considered the two loop contribution
associated with the interaction of the radiation scalar field with the quarks
which diagram is shown in figure \ref{fig3}. 

It followed that the analytic form of the Feynman diagram became very similar
to the previous one
\begin{equation}
\Gamma_{Y}^{(2)}[\phi]=V^{(D)}2N\int\frac{dp^{D}dq^{D}}{i^{2}\text{ }%
(2\pi)^{2D}}\frac{p^{2}-\frac{q^{2}}{4}+\phi^{2}}{q^{2}((p+\frac q2)^{2}%
-\phi^{2})((p-\frac q2)^{2}-\phi^{2})},
\end{equation}
and it was calculated also in a close manner to find for the action density
and potential associated to this term
\begin{align}
\gamma_{Y}^{(2)}[\phi]  & =-v_{Y}^{(2)}[\phi]\nonumber\\
& =\frac{4N}{\text{ }(2\pi)^{8-4\epsilon}}\frac{A(\epsilon)\pi^{4-2\epsilon}%
}{\epsilon^{2}}\phi^{4}(\frac{\phi^{2}}{\mu^{2}})^{-2\epsilon}\nonumber\\
& -\frac N{\text{ }(2\pi)^{8-4\epsilon}}\pi^{4-2\epsilon}(\Gamma
(\epsilon-1))^{2}\phi^{4}(\frac{\phi^{2}}{\mu^{2}})^{-2\epsilon}.
\end{align}
Subtracting the divergent terms in the Laurent expansion with respect to the
parameter $\epsilon$, gave the following finite contribution.
\begin{align}
v_{Y}^{(2)}[\phi]=-0.0000601522\text{ }m^{4}\left(  332.744-123.833Log[\frac
{m^{2}}{\mu^{2}}]+12.0 \, Log[\frac{m^{2}}{\mu^{2}}]^{2}\right) .
\end{align}
\newline
\begin{figure}[h]
\centering
\includegraphics[width=11cm]{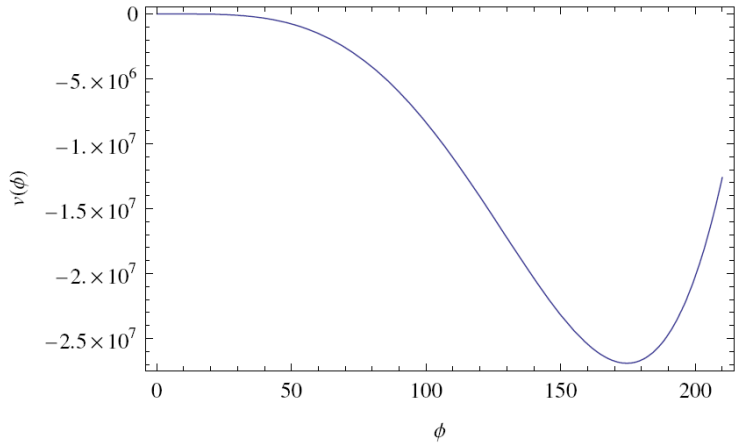} \caption{ The effective potential for
the ''mass'' field $\phi$ for the value of the scale $\mu$ determining that
the value of $\phi$ taken at the minimum potential, is approximately equal to
the top quark mass $m_{top}~175\,\,GeV$. }%
\label{fig4}%
\end{figure} 
Then, in reference \cite{cabo}, it was further considered the study of the
potential and its minimum as a function of the parameters. Specifically, the
coupling and renormalization scale parameter $\mu$ were chosen as satisfying
the one loop RG expression.
\begin{equation}
g_{0}(\mu,\Lambda_{QCD})=2\sqrt{\frac27}\pi\sqrt{\frac1{Log(\frac\mu
{\Lambda_{QCD}})}},
\end{equation}
and the value $\Lambda_{QCD}$ = 217 MeV was selected. Next, $\mu$ was changed
up to fix the mean scalar field $\phi$ to a value determining the top quark
mass: 174.599 GeV. 
The resulting curve for the scalar field potential is shown in
figure \ref{fig4}. An interesting conclusion implied by this plot was that the
second derivative at the potential minimum is $m_{\phi}=126.76\,\,GeV$, which
is close to the recently measured value of the Higgs mass. At the year 2011 in
which the paper \cite{cabo}  appeared, the Higgs mass was still unknown.
However, at that time the mass value $114$ $GeV$, indicated by some low
statistic data measured at ALEPH experiment at CERN, led to suspect the link
of the result with the Higgs mass.  However, it is needed to remark that in
the simplified model considered in \cite{cabo}, the scale $\mu$ , which
allowed the top mass fixation lies within the non perturbative infrared
region: $\mu=0.498\,\,GeV$, which determines a coupling value $\alpha
=\frac{g_{0}^{2}}{4\pi}=1.077$. Therefore, the result about the possibility of
basing the SM in the "second minimum" remained inconclusive at that time.
Therefore, the possibility of fixing the observed value of the top mass in the
scheme, and with it, the $126.76\,\,GeV$ estimation of the Higgs particle mass
remained open. 
\begin{figure}[h]
\centering
\includegraphics[width=7cm]{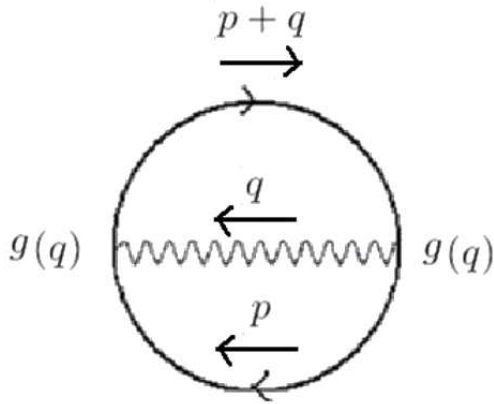}
\caption{The two loop contribution depending on the momentum dependent running coupling. }
\label{fig5}
\end{figure}

\subsection*{The proposal}

As it was mentioned, a main objective of the present letter, in addition to
reviewing the ideas of the paper \cite{cabo}, is to identify and advance a
possibility for the normal strength of the strong coupling to becomes able in
justifying the physical values for the Top and Higgs masses in the SM. As it
was mentioned, the central limitation of the results discussed in reference
\cite{cabo}, was the fact that a slightly high value of strong coupling was
required to  fix the Top and Higgs masses. The coupling values employed for
the evaluations were assumed as constants. Therefore, the idea comes to the
mind about that the running with momenta  strong coupling should increase at
low momenta. Henceforth, the possibility appears that using a momenta
dependent coupling in the two loop evaluation of the closed fermion loop
contracted with the gluon propagator, could furnish similar results for the
Higgs potential that the ones obtained for the relatively high constant
coupling employed in reference \cite{cabo}. The two loop contribution
associated to the quark gluon interaction in the background of the scalar
field, but including a transferred gluon momentum $q$ dependence of the strong
coupling is shown in figure \ref{fig5}. The calculation of the expression in
which the coupling is running becomes more involved, since the evaluations of
the two loop integrals should be done numerically due to the momentum
dependence of the $g(p)$. It is clear that the first integration over the
momentum $p$ can be done, but it will include one loop infinities. Thus the
best way of proceeding seems to be first constructing the renormalized form of
the model, for further evaluating the finite part of the diagram under
consideration. Afterwards, the best form of inputting the running coupling
should more easily discussed.

The investigation of this question will be considered elsewhere. If the
conclusions become positive ones, the measured Top and Higgs masses might be
defined by a symmetry breaking based in the \textit{second
minimum} of the Higgs potential, in place than on the Higgs
classical minimum.

\acknowledgements*
A.C. would like to acknowledge a helpful discussion with 
Prof. Masud Chaichian in which he underlined  the motivating possibility that
 the running coupling constant in place of a constant coupling.

\end{document}